\documentclass[epj]{svjour}
\usepackage{latexsym}
\usepackage{graphics}
\usepackage{amssymb}
\usepackage{amsmath}
\usepackage[switch,mathlines]{lineno}
\begin{document}
\title{Nuclear resonance fluorescence of $^{208}$Pb in heavy-ion colliders}
\author{Uliana~Dmitrieva\inst{1,2} \and Igor~Pshenichnov\inst{1,2}
}                     
%
\mail{pshenich@inr.ru (I.~Pshenichnov)}
\institute{Moscow Institute of Physics and Technology, 9 Institutskiy per., Dolgoprudny,
141700 Moscow Region, Russia \and Institute for Nuclear Research, Russian Academy of Sciences, 60-letia Okatyabrya, 7a, 
117312 Moscow, Russia}
\date{Received: date / Revised version: date}
%

\abstract{
In ultraperipheral collisions (UPC) of nuclei the impact of Lorentz-contracted electromagnetic fields of collision partners leads to their excitations. In case of heavy nuclei the emission of neutrons is a main deexcitation channel and forward neutrons emitted in UPC were detected at the Relativistic Heavy-Ion Collider (RHIC) and at the Large Hadron Collider (LHC) by means of Zero Degree Calorimeters. However, the excitation of low-lying discrete nuclear states is also possible in UPC below the neutron separation energy. In this work by means of the Weizsacker-Williams method the data on nuclear resonance fluorescence (NRF) induced by real photons in $^{208}$Pb are used to model the excitations of discrete levels in colliding nuclei. Due to Lorentz boosts one can expect that deexcitation photons with energies up to 40~GeV and 300~GeV are emitted in very forward direction, respectively, at the LHC and at the Future Circular Collider (FCC-hh). Energy, rapidity and angular distributions of such photons are calculated in the laboratory system, which can be used for monitoring of collider luminosity or triggering particle production in UPC. 
} 
\maketitle

\section{Introduction}
\label{intro}
Since the 1950's electromagnetic (Coulomb) excitation of nuclei by charged particles has been used to probe nuclear structure~\cite{McGowan1958,Stelson1963}. Such studies were mostly focused on the properties of nuclear states excited by electromagnetic fields because the properties of electromagnetic interaction were already well understood.  A variety of nuclear and particle physics phenomena can be be studied in ultraperipheral collisions (UPC) of nuclei, where the collision impact parameter $b$ exceeds the sum of the radii, $R_1$ and $R_2$, of colliding nuclei: $b>R_1+R_2$~\cite{Bertulani2005}. In particular, properties of exotic nuclei far away from the valley of stability have been investigated in UPC with stable nuclei~\cite{Aumann2005,Gade2008}. Electromagnetic excitation of collective states like giant dipole resonances (GDR) in nuclei typically results in their electromagnetic dissociation (EMD) with emission of neutrons and protons~\cite{Bertulani1988}. The excitation and decay of GDR is the most important UPC channel and multiphonon GDR excitations were successfully studied in UPC~\cite{Aumann1998,Bertulani1999}. More recent studies include the investigations of EMD of nuclei with a neutron skin~\cite{Klimkiewicz2007,Bracco2011}. Excited nuclear states below the GDR region are known as pygmy resonances~\cite{Savran2013,Tonchev2015}, and their coupling to giant resonances have been also investigated~\cite{Brady2016}.  

A lot of data on $^{197}$Au--$^{197}$Au, $^{238}$U--$^{238}$U and $^{208}$Pb--$^{208}$Pb collisions have been collected, respectively, at the Relativistic Heavy-Ion Collider (RHIC) at Brookhaven National Laboratory (BNL) and at the Large Hadron Collider (LHC) at the European Organization for Nuclear Research (CERN). In addition to hadronic interactions of nuclei with overlap of their nuclear densities, the research programs at RHIC and the LHC include studies of UPC~\cite{Baltz2008}. As demonstrated by measurements and calculations~\cite{Fischer2014}, the uranium beam lifetime at RHIC is strongly influenced by the bound-free ${\rm e}^+{\rm e}^--$ pair production (BFPP) and EMD of $^{238}$U circulating in RHIC.  At the LHC both processes are responsible for the production of secondary ions from $^{208}$Pb$^{82+}$ which can potentially impact superconducting magnets and cause their quenching~\cite{Bruce2009,Bruce2010}. The cross sections of the most important channels of EMD of $^{208}$Pb  calculated by RELDIS Monte Carlo model~\cite{Pshenichnov2011} agree well with the cross sections of forward neutron emission measured by the ALICE experiment~\cite{Abelev2012} at the LHC. As shown in these publications, the EMD of ultrarelativistic nuclei with emission of nucleons is presently well understood. The detection of  forward neutrons by Zero Degree Calorimeters (ZDC)~\cite{Puddu2007,Oppedisano2009,Cortese2019} helps to monitor the collider luminosity~\cite{Pshenichnov2011} and can be also used as a trigger of UPC events with particle production~\cite{Baltz2008}. 

However, much less attention has been payed so far to low-energy excitations of ultrarelativistic nuclei in UPC below the nucleon separation energy and thus resulting exclusively to emission of few MeV photons in the nucleus rest frame. It is expected that due to the Lorentz boost energetic forward photons will be emitted in the laboratory system. This process is similar to the nuclear resonance fluorescence (NRF), which has been studied intensively in experiments with real photons~\cite{Mohr1999,Shizuma2011,Savran2015} and proposed, in particular, for non-destructive assays of spent nuclear fuel~\cite{Hayakawa2010}. The properties of numerous discrete excited nuclear states in $^{204,206,207,208}$Pb below the neutron emission threshold were studied by detecting NRF photons after the impact of bremmstrahlung radiation with the endpoint energy of 6.75~MeV~\cite{Enders2003}. While the interest in detecting forward photons emitted by nuclear spectators in nucleus-nucleus collisions has been clearly expressed~\cite{Norbeck2012}, very forward photons at the LHC were detected exclusively by the LHCf experiment in pp and p--Pb collisions~\cite{Tiberio2015,Menjo2012,Adriani2011,Adriani2012}. To the best of our knowledge there are only two theory papers~\cite{Korotkikh2002,Kharlov2004} devoted to the excitation of discrete nuclear levels in UPC at the LHC and the respective photon emission. In the first work~\cite{Korotkikh2002} the process of ${\rm e}^+{\rm e}^--$ pair production in fusion of virtual photons in UPC of $^{40}$Ca nuclei followed by the excitation of discrete levels in $^{40}$Ca by the electron or positron with a subsequent photon emission has been investigated. 
In the second work~\cite{Kharlov2004} the process of production of a meson in $\gamma\gamma$-fusion 
with simultaneous excitation of discrete nuclear levels $^{16}$O$^\star$~(2$^+$, 6.92~MeV) and   $^{208}$Pb$^\star$~(3$^-$, 2.62~MeV), respectively, in $^{16}$O--$^{16}$O and $^{208}$Pb--$^{208}$Pb UPC at the LHC has been studied.

In the present work direct excitation by Weizsacker-Williams photons of discrete levels of $^{208}$Pb in their UPC at the LHC and at the Future Circular Collider (FCC-hh)~\cite{Bogomyagkov2016,Benedikt2016} is considered. All known excited states below the nucleon separation energy of $^{208}$Pb are taken into account. Their decays are only possible by emission of photons, and the properties of this radiation are investigated.
\section{Total NRF cross section for ultraperipheral collisions of $^{208}$Pb at the LHC and FCC-hh}
\label{NRF_cross_section}
\begin{figure}
\resizebox{0.5\textwidth}{!}{
  \includegraphics{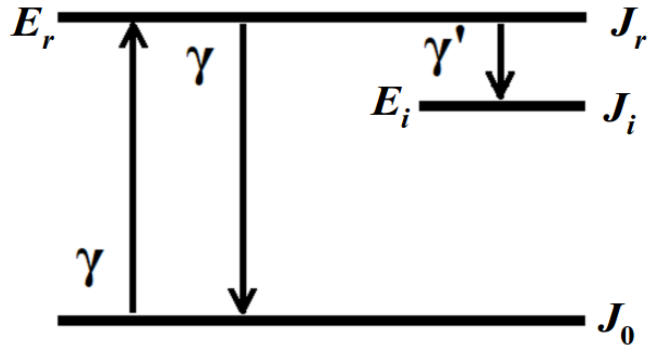}
}
\caption{Nuclear resonance fluorescence: (1) the excitation of an isolated nuclear state with the energy $E_r$, spin $J_r$, total width $\Gamma_r$ by a photon $\gamma$; (2) subsequent decays either to an intermediate state ($E_i$, $J_i$) with partial width $\Gamma_i$ by emitting a photon $\gamma'$ or back to the ground state ($J_0$) with partial width $\Gamma_0$ by emitting the photon $\gamma$.} 
\label{fig:NRF_scheme}
\end{figure}

The phenomenon of NRF consists in (1) the excitation of an isolated nuclear state with the energy $E_r$, spin $J_r$, total width $\Gamma_r$ by the absorption of a photon; and (2) subsequent decays either to an intermediate state $i$ with $E_i$, $J_i$ or directly to the ground state with the spin $J_0$, see Fig.~\ref{fig:NRF_scheme}. 
In the case of a relatively high initial excitation, its total width $\Gamma_r$ is given by the sum of partial widths $\Gamma_r=\Gamma_0 + \sum_i \Gamma_i$ corresponding to decays to the ground state and intermediate states. However, for the purposes of this work, it is sufficient to assume that the direct decay to the ground state dominates, $\Gamma_r \sim \Gamma_0$, and cascade transitions to the ground state through intermediate states can be also represented by this direct transition. Indeed, cascade and direct NRF events remain indistinguishable,  provided that the sum of the energies of cascade photons in each event is measured by a forward electromagnetic calorimeter and is thus close to $E_r$. Then, for the direct transition to the ground state ($J_r \rightarrow J_0$), the NRF cross section is described by the Breit-Wigner formula:
\begin{equation}
 \sigma(E_\gamma)=\frac{\pi}{2}\left(\frac{\hbar c}{E_r}\right)^2 g\frac{\Gamma_0^2}{(E_\gamma-E_r)^2+\Gamma_r^2/4} \ .
\end{equation}
Here $E_\gamma$ is the photon energy, $g=(2J_r+1)/(2J_0+1)$ is the spin statistical factor, $\Gamma_0$ is the partial width of the decay back to the ground state. The width $\Gamma_r$ is typically of few eV.

The integral cross section for the considered excited state with the energy $E_r$ is calculated as:
\begin{equation}
 I(E_r)=\int {\rm d}E_\gamma \sigma(E_\gamma) = \pi^2 \left(\frac{\hbar c}{E_r}\right)^2 g \frac{\Gamma_0^2}{\Gamma_r}=\frac{\pi}{2}\Gamma_r\sigma_{\rm max}(E_r) \ ,
\end{equation}
where the maximum cross section value at $E_r$ is: 
\begin{equation}
 \sigma_{\rm max}(E_r)=2\pi \left(\frac{\hbar c}{E_r}\right)^2 g \frac{\Gamma_0^2}{\Gamma_r^2} \ .
\end{equation}

The calculation of the NRF cross section $\sigma_{\rm NRF} (E_r)$ in ultraperipheral collisions for a given resonant state $E_r$ is straightforward. In contrast to the applications of the Weizsacker-Williams method at lower collision energies, where the contribution of each electromagnetic multipole is treated separately in the spectrum: 
\begin{equation}
\begin{aligned}
n_{\rm WW}^{E2}(E_\gamma)\gg n_{\rm WW}^{E1}(E_\gamma)\gg n_{\rm WW}^{M1}(E_\gamma)\, ,   
\end{aligned} 
\end{equation}
see, e.g.,~\cite{Cerutti2017}, the sum over different multipolarities can be easily obtained at ultrarelativistic energies. As discussed~\cite{Bertulani1988}, 
\begin{equation}
\begin{aligned}
n_{\rm WW}^{E2}(E_\gamma)\approx n_{\rm WW}^{E1}(E_\gamma)\approx n_{\rm WW}^{M1}(E_\gamma) = n_{\rm WW}(E_\gamma)
\end{aligned} 
\end{equation}
when $\gamma \gg 10, \beta \rightarrow 1$. Therefore, the multipole expansion can be replaced by the sum of the respective partial cross sections  $\sigma(E_\gamma)=\sum_{\pi l}\sigma^{\pi l}(E_\gamma)$ multiplied by $n_{\rm WW}(E_\gamma)$.
Moreover, the variations of the WW spectrum $n_{\rm WW}(E_\gamma)$~\cite{Bertulani2005} within a very small resonance width $\Gamma_r$ can be safely neglected. As a result, $n_{\rm WW}(E_\gamma)$ can be represented by its value at $E_r$ to calculate the respective integral:  
\begin{equation}
\label{NRF_f}
\begin{aligned}
\sigma_{\rm NRF} (E_r) = \int {\rm d}E_\gamma \sigma(E_\gamma) n_{\rm WW}(E_\gamma) = n_{\rm WW}(E_r)I(E_r)\\
 =  \frac{\pi}{2} n_{\rm WW}(E_r)\Gamma_r\sigma_{\rm max}(E_r)\ .
\end{aligned}
\end{equation}

\begin{figure}[htb]
\resizebox{0.5\textwidth}{!}{%
  \includegraphics{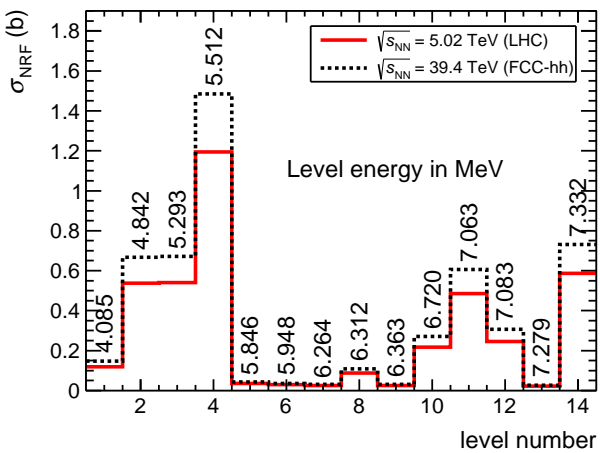}
}
\caption{NRF cross sections for ultraperipheral $^{208}$Pb--$^{208}$Pb collisions at the LHC and FCC-hh, respectively, at $\sqrt{s_{\rm NN}}=5.02$ TeV (solid histogram) and at $\sqrt{s_{\rm NN}}=39.4$~TeV (dashed histogram).}
\label{fig:NRF}
\end{figure}

In Table~\ref{tab:NRF} the values of energy $E_r$, width $\Gamma_r$ and $\sigma_{\rm max}$ for excitations of discrete levels in $^{208}$Pb below the neutron emission threshold in the energy range from 4 to 8~MeV~\cite{Yavas2005} are given. In total, 14 levels in $^{208}$Pb are taken into account. Their  parameters were used to calculate NRF cross sections for each of these exited states in ultraperipheral $^{208}$Pb--$^{208}$Pb collisions according to Eq.~(\ref{NRF_f}). The resulting cross sections for collisions at the LHC and FCC-hh at $\sqrt{s_{\rm NN}}=5.02$ and 39.4~TeV, respectively, are given in Table~\ref{tab:NRF}. The values of the NRF cross sections were also obtained numerically by means of CERN ROOT as integrals  of $\sigma(E_\gamma)\times n_{\rm WW}(E_\gamma)$ individually for each peak, instead of using Eq.~(\ref{NRF_f}). The largest differences between the results of the numerical integration and those of Eq.~(\ref{NRF_f}) were found for small $E_r$, as well as for the peaks with largest $\Gamma_r$. However, in all cases the relative differences were less than $7\times 10^{-6}$. This demonstrates the applicability of Eq.~(\ref{NRF_f}) for calculating NRF cross sections in collisions of ultrarelativistic nuclei. 
\begin{table*}[htb]
\caption{Characteristics of excited states in $^{208}$Pb below neutron emission threshold and corresponding NRF cross sections for ultraperipheral $^{208}$Pb--$^{208}$Pb collisions at the LHC and FCC-hh at $\sqrt{s_{\rm NN}}=5.02$ and 39.4~TeV, respectively.}
\label{tab:NRF}      
\begin{tabular}{ccccccc}
\hline
Level  & Energy    & Width      & Spin,        & Photoabsorption cross            & \multicolumn{2}{c}{NRF cross section} \\
number & $E_r$     & $\Gamma_r$ & parity       & section $\sigma_{\rm max}(E_r)$  & \multicolumn{2}{c}{$\sigma_{\rm NRF}(E_r)$ }  \\
       & (MeV)     & (eV)       & $J_r^{\pi}$  &  (b)                             & \multicolumn{2}{c}{(b)}  \\
\cline{6-7}       
       &           &            &              &                         & $\sqrt{s_{\rm NN}}=5.02$~TeV & $\sqrt{s_{\rm NN}}=39.4$~TeV \\
\hline
1 & 4.085 & 0.783 & 2+ & 730.6 & 0.118 & 0.147\\
2 & 4.8422 & 9.972 & 1-- & 312.0 & 0.538 & 0.668\\
3 & 5.2926 & 13.16 & 1-- & 261.2 & 0.541 & 0.672\\
4 & 5.5122 & 32.91 & 1-- & 240.8 & 1.194 & 1.485\\
5 & 5.8461 & 1.154 & 1+ & 214.1 & 0.035 & 0.044\\
6 & 5.9480 & 1.012 & 1-- & 206.8 & 0.029 & 0.036\\
7 & 6.2640 & 1.012 & 1-- & 186.5 & 0.025 & 0.031\\
8 & 6.3117 & 3.656 & 1-- & 183.6 & 0.088 & 0.109\\
9 & 6.3628 & 1.044 & 1-- & 180.7 & 0.024 & 0.030\\
10 & 6.7205 & 10.97 & 1-- & 162.0 & 0.217 & 0.270\\
11 & 7.0635 & 28.61 & 1-- & 146.6 & 0.486 & 0.606\\
12 & 7.0834 & 14.62 & 1-- & 145.8 & 0.246 & 0.307\\
13 & 7.2789 & 1.4 & 1+ & 138.1 & 0.022 & 0.027\\
14 & 7.3325 & 38.71 & 1-- & 136.1 & 0.587 & 0.732\\
\hline
Sum & & & & & 4.15 & 5.16\\
\hline
\end{tabular}
\end{table*}
The calculated NRF cross sections are also shown in Fig.~\ref{fig:NRF}. As can be seen from this figure, the contribution of low-lying levels is enhanced because the Weizsacker-Williams spectrum $n_{WW}(E_{\gamma})$ changes with equivalent photon energy $E_{\gamma}$ approximately as $1/E_{\gamma}$. As a result, the contribution of first four levels ($\sim 58$\%) dominates at both collision energies. The sum of NRF cross sections for all considered levels amounts to 4.15 b and to 5.16 b, respectively, at $\sqrt{s_{\rm NN}}=5.02$ and 39.4~TeV, see Table~\ref{tab:NRF}.
\section{Angular, energy and rapidity distributions of photons in the laboratory system}\label{sec:distributions}

There are two features of ultraperipehral collisions of ultrarelativistic nuclei which simplify the calculation of the angular distribution of photons emitted by these nuclei. Firstly, because of various orientations of the reaction plane, spin states of excited nuclei are not aligned and one can assume that photons from nuclear de-excitation are emitted isotropically in the rest frame of emitting nucleus. Secondly, because of negligible changes of the total energy of a nucleus $E_A=\gamma M_A$ due to the absorption of an equivalent photon, one can safely assume that this nucleus propagates further along the beam direction with initial velocity. This is because of the energy of equivalent photons is restricted by $E_{max}\sim \gamma/R$ and this value represents only a negligible fraction of the total energy of the nucleus. For example, for heavy nuclei $r=E_{max}/E_A\approx 1/R M_A\sim 10^{-4}$.  In this estimation, as well as in the following text natural units are used: $\hbar =c=1$ .  

One can consider an excited nucleus with the Lorentz-factor $\gamma$ which emits a photon with the energy $E_r$ and momentum $p_r = E_r$ in its rest frame. In this reference frame the angle between the photon momentum and the beam direction can be denoted as $\theta_r$. In the laboratory frame the photon energy $E$, its longitudinal $p_L$ and transverse $p_T$ momentum components with respect to the beam direction are given by the corresponding Lorentz boost:
\begin{equation}
\left\{ 
\begin{array}{l}
   E = \gamma(E_r + \beta p_r \cos{\theta_r})\\
   p_L = \gamma(\beta E_r +p_r \cos{\theta_r})\\
   p_T = p_r \sin{\theta_r}\\
\end{array}
\right.
\label{eq:Lorentz-boost}
\end{equation}
Thus, the angle $\theta$ of photon emission in the laboratory frame can be calculated from the relation: 
\begin{equation}
\tan{\theta} = \frac{p_T}{p_L} = \frac{\sin{\theta_r}}{\gamma(\beta + \cos{\theta_r})} \ .
\end{equation}
Because of $\gamma \gg 1$ and¨ $\beta\approx 1$ this is reduced to 
\begin{equation}
\tan{\theta} = \frac{1}{\gamma} \tan{\frac{\theta_r}{2}} \ .
\end{equation}
This indicates that the angular distribution of photons emitted in the nucleus rest frame is extremely contracted by factor of $1/\gamma$ in the laboratory system in the case of ultrarelativistic nuclei. This is a well-known projector effect which makes insignificant the differences in angular distributions of emitted photons for individual levels with various $J_r^{\pi}$. In other words, all these photons can be detected by a quite compact forward detector placed far from the interaction point.

\begin{figure}[htb]
\resizebox{0.5\textwidth}{!}{%
  \includegraphics{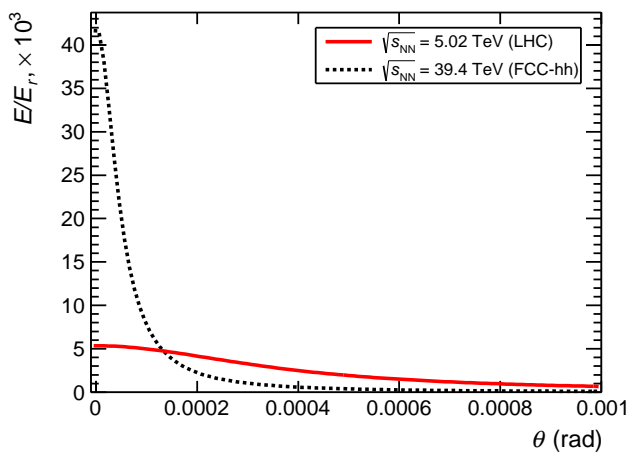}
}
\caption{The ratio $E/E_r$ (note the factor of $10^{3}$) as a function of photon emission angle $\theta$ in the laboratory frame for $^{208}$Pb beams colliding at the LHC at $\sqrt{s_{\rm NN}}=5.02$~TeV and at the FCC-hh at $\sqrt{s_{\rm NN}}=39.4$~TeV.}
\label{fig:Dopler}
\end{figure}

The photon energy $E$ in the laboratory system depends on the angle of emission $\theta_r$ in the nucleus rest frame. With the condition $\gamma \gg 1$, $\beta\approx 1$ the dependence of $E$ on emission angle $\theta$ in the laboratory system can be simplified:
\begin{equation}
\begin{gathered}
   E= \gamma(E_r + \beta p_r \cos{\theta_r})\approx \gamma E_r(1+\cos\theta_r) = 2\gamma E_r \cos^2\frac{\theta_r}{2}    \\
     =\frac{2\gamma E_r}{1+\tan^2\frac{\theta_r}{2}} =\frac{2\gamma E_r}{1+\gamma^2 \tan^2{\theta}} \ ,   
\end{gathered}
\label{eq:ETheta}
\end{equation}
where $E_r$ is the photon energy in the nucleus rest frame. It is convenient to express the ratio $E/E_r$ as a function of $\theta$, as shown in Fig.~\ref{fig:Dopler} for $^{208}$Pb beams at the LHC and FCC-hh, because this dependence is valid for NRF from any level in $^{208}$Pb at a specific $\gamma$.
In the case of very forward photon emission at $\theta=0$, the photon energy reaches its maximum value of $E_{max}=2\gamma E_r$. This means that energy of NRF photons in the laboratory frame extends up to $\sim 40$~GeV and $\sim 300$~GeV, respectively, for ultraperipheral $^{208}$Pb--$^{208}$Pb collisions at $\sqrt{s_{\rm NN}}=5.02$~TeV and 39.4~TeV at the LHC and FCC-hh because of huge Lorentz-factors: $\gamma= 2.7\times 10^3$ and $2.1\times 10^4$.

As explained above, an isotropic distribution of photon emission in the rest frame of $^{208}$Pb can be safely assumed: ${\rm d}N/{\rm d}\Omega_r =1/4\pi$, or, after the integration over azimuthal angle, ${\rm d}N/{\rm d}\theta_r=-\sin\theta_r/2$. From the condition that the number of photons emitted within a given angle element in both reference systems is an invariant, the angular distribution in the laboratory system can be obtained: 
\begin{equation}
\frac{{\rm d}N}{{\rm d}\theta} = - \frac{\sin\theta_r}{2} \frac{{\rm d}\theta_r}{{\rm d}\theta} \ .
\label{eq:invariant}
\end{equation}
Then, by calculating
\begin{equation}
 \begin{gathered}
\sin\theta_r = \frac{2\tan\theta_r/2}{1+\tan^2\theta_r/2} = \frac{2\gamma\tan\theta}{1+\gamma^2\tan^2\theta} \\
=\frac{2\gamma\sin\theta}{(1+\gamma^2\tan^2\theta)\cos\theta}
\end{gathered}
\end{equation}
and 
\begin{equation}
\frac{{\rm d}\theta_r}{{\rm d}\theta}=2 \frac{{\rm d}(\arctan(\gamma\tan\theta))}{{\rm d}\theta}=\frac{2\gamma}{(1+\gamma^2\tan^2\theta)\cos^2\theta}\ ,
\end{equation}
one finally obtains for the interval of $\theta$ from 0 to $\pi/2$:
\begin{equation}
\frac{{\rm d}N}{{\rm d}\theta} = -\frac{2\gamma^2\sin\theta}{(1+\gamma^2\tan^2\theta)^2\cos^3\theta} = -\frac{2\gamma^2\sin\theta\cos\theta}{(1+(\gamma^2-1)\sin^2\theta)^2} \  .
\end{equation}
Alternatively, for the same interval of $\theta$ the angular distribution can be expressed only via $\tan\theta$:
\begin{equation}
\frac{{\rm d}N}{{\rm d}\theta}=-\frac{2\gamma^2\tan\theta(1+\tan^2\theta)}{(1+\gamma^2\tan^2\theta)^2} \ .
\label{eq:thetatangent}
\end{equation}

The distribution of emitted photons can be expressed in terms of pseudorapidity~$\eta = -\ln(\tan\theta/2)$ instead of angle $\theta$. With 
\begin{equation}
\frac{{\rm d}\theta}{{\rm d}\eta}=-\frac{2{\rm e}^{\eta}}{{\rm e}^{2\eta}+1}=-\frac{1}{\cosh\eta} \ ,
\end{equation}
and
\begin{equation}
\tan\theta=\frac{2\tan\theta/2}{1-\tan^2\theta/2}=\frac{2{\rm e}^{-\eta}}{1-{\rm e}^{-2\eta}}=\frac{1}{\sinh\eta} \ ,
\label{eq:TanEta}
\end{equation}
using Eq.(\ref{eq:thetatangent}) the pseudorapidity distribution is obtained:
 \begin{equation}
 \begin{gathered}
\frac{{\rm d}N}{{\rm d}\eta} = 2\gamma^2 \frac{1+\sinh^{-2}\eta}{(1+\gamma^2\sinh^{-2}\eta)^2 \sinh\eta\cosh\eta}\\ 
=  \frac{2\gamma^2\sinh\eta\cosh\eta}{(\gamma^2+\sinh^2\eta)^2} \ .
\end{gathered}
 \end{equation}

\begin{figure}[ht]
\resizebox{0.5\textwidth}{!}{%
  \includegraphics{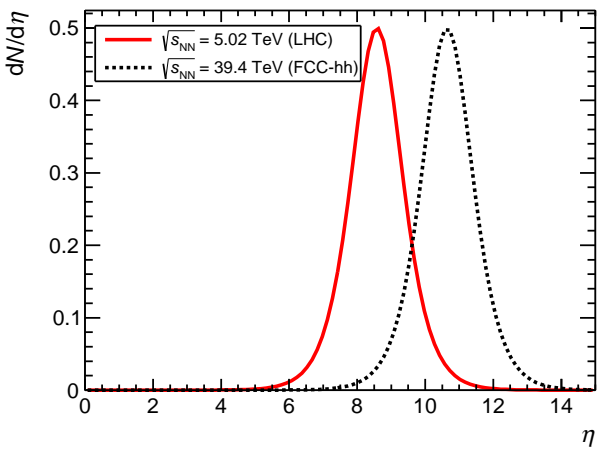}
}
\caption{Pseudorapidity distribution of NRF photons calculated for ultraperipheral $^{208}$Pb--$^{208}$Pb collisions at the LHC at $\sqrt{s_{\rm NN}}=5.02$~TeV and at the FCC-hh at $\sqrt{s_{\rm NN}}=39.4$~TeV.}
\label{fig:dNdeta}
\end{figure}

\begin{figure}[htb]
\resizebox{0.5\textwidth}{!}{%
  \includegraphics{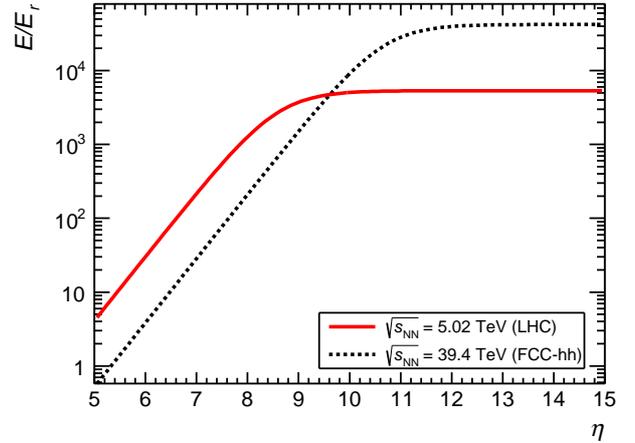}
}
\caption{The ratio $E/E_r$ as a function of pseudorapidity $\eta$ for $^{208}$Pb beams colliding at the LHC at $\sqrt{s_{\rm NN}}=5.02$~TeV and at the FCC-hh at $\sqrt{s_{\rm NN}}=39.4$~TeV.}
\label{fig:EEreta}
\end{figure}

 Such pseudorapidity distributions are shown in Fig.~\ref{fig:dNdeta} for ultraperipheral $^{208}$Pb--$^{208}$Pb collisions at the LHC and FCC-hh. Since they are calculated for massless photons, ${\rm d}N/{\rm d}\eta \equiv {\rm d}N/{\rm d}y$. As expected, the distributions have distinct maxima at the very forward direction corresponding to the beam rapidities of $y_{\rm beam}=8.6$ and 10.7 for collisions at $\sqrt{s_{NN}}=5.02$~TeV and 39.4~TeV, respectively. While these distributions are quite wide and extend for six units of pseudorapidity, NRF photons interact only with forward detectors. For example, it is expected that in ALICE experiment such photons hit only Zero Degree Calorimeters~\cite{Oppedisano2009,Cortese2019} ($|\eta|>8.8$) and ALICE Diffractive detector (AD)~\cite{Tello2017}. In particular, one side of AD, which covers $-7<\eta < -4.9$, may be affected by NRF photons. However, photons entering AD are rather soft ($E<1$~GeV), as follows from the dependence of their energy on pseudorapidity:
 \begin{equation}
 E=\frac{2\gamma E_r}{1+\gamma^2\sinh^{-2}\eta} \ ,
 \end{equation}
obtained from Eqs.~(\ref{eq:ETheta}) and (\ref{eq:TanEta}). $E/E_r$ as functions of $\eta$ are shown in Fig.~\ref{fig:EEreta} for $^{208}$Pb--$^{208}$Pb UPC at the LHC and FCC-hh.

\section{Impact of level lifetime}\label{sec:lifetime}
Since the considered low-lying levels in $^{208}$Pb are characterized by finite lifetimes, their de-excitations do not take place exactly at the interaction point, but rather away from it. A typical distance between a point of a $^{208}$Pb--$^{208}$Pb collision and a photon emission point can be estimated for the level with the largest NRF cross section. Its energy is $E_r=5.5122$~MeV, and the width is $\Gamma_r=32.91$~eV. Therefore, the lifetime of the excited state in the nucleus rest frame is $\tau = \hbar / \Gamma_r = 2\cdot 10^{-17}$~s, while at the LHC in the laboratory frame it is $t=\gamma \cdot \tau \approx 5.35\cdot 10^{-14}$~s for $\gamma= 2.7\times 10^3$. During time $t$ an ultrarelativistic $^{208}$Pb moves only for a small distance of $l=c\cdot t \approx 16$~$\mu$m at the LHC.  For the narrowest level with $E_r=4.085$~MeV, $\Gamma_r=0.783$~eV this distance is 42 times longer, $l\approx 0.67$~mm, but still  microscopic. The estimations of $l$ at the FCC-hh are naturally about eight times larger following the ratio of the corresponding Lorentz factors. It is $l\approx 5.2$~mm for the narrowest level, but still on a millimeter scale.      

It can be noted that $l$ is comparable to the typical transverse RMS widths of LHC beams delivered to interaction points, in particular, for ALICE experiment (from 15 to 150 $\mu$m)~\cite{Abelev2014}.  However, the RMS of the longitudinal size of the collision region in this experiment is much larger than $l$,  about 6~cm~\cite{Abelev2014}. In general, the estimated distance $l$ is smaller than the uncertainty in determining the longitudinal vertex position in the ALICE experiment. All of this suggests, that NRF photons are emitted in the very centers of the detectors installed at the LHC. Therefore, keeping in mind a possibility of their detection, NRF photons can be proposed for improving the determination of vertex position.  
\section{Conclusions}\label{sec:conclusions}
We considered the process of nuclear resonance fluorescence (NRF) induced by Weizsacker-Williams photons in ultraperipheral collisions of ultrarelativistic $^{208}$Pb at the LHC and FCC-hh. The interest to such a phenomenon is truly interdisciplinary, because the nuclear structure of $^{208}$Pb, which is usually studied in low-energy physics, can be also probed in high-energy physics experiments. A more sophisticated approach to probe the levels in $^{208}$Pb by irradiating beam nuclei at the LHC with a free electron laser has been proposed in Ref.~\cite{Yavas2005} to make an intense source of 1~MeV$<E_{\gamma}<400$~MeV photons. The status of the photon-nucleus collider proposed in Ref.~\cite{Yavas2005} is unclear, but it certainly presumes very high costs of the installation and operation. In contrast, the registration of NRF photons considered in the present work requires only the creation of traditional electromagnetic calorimeters, possibly with advanced transverse segmentation suitable for measurements of a sharply-forward angular distributions of these photons.    

As shown in the present work, photons with energy up to 40~GeV are frequently emitted in the forward direction because of the de-excitation of discrete levels in $^{208}$Pb following their ultraperipheral collisions at the LHC. This can be also considered as a source of monochromatic photons of higher energies compared to Ref.~\cite{Yavas2005}. Such photon emission is also expected at the FCC-hh, with photon energies up to 300~GeV. The total cross section of photon emission is estimated as large as 4.15~b and 5.16~b at the LHC and FCC-hh, respectively. Convenient energy, angular and pseudorapidity distributions of NRF photons have been obtained analytically. Similarly to the studies~\cite{Korotkikh2002,Kharlov2004} of other processes of photon emission due to the de-excitation of discrete levels in ultraperipheral nucleus-nucleus collisions such distributions can be used for estimating the impact on various detectors or monitoring collision rate.  

Finally, it can be noted that in the LHCf experiment~ \cite{Adriani2012} forward photons are detected starting from 50~GeV. Since the energy of NRF photons is proportional to the energy of the nuclear beam, further increase in beam energy in future LHC runs will allow to detect NRF photons using the existing LHCf equipment.

 \bibliographystyle{epj}
 \bibliography{Dmitrieva_Pshenichnov_NRF}

\end{document}